# Physicochemical Perturbations of Phase Equilibriums

*Vladimir Kh. Dobruskin*

11 Pashosh St., Beer-Yacov 30700, Israel

Tel: (972)-506-864 642; Fax: (972)-77-531 34 48; E-mail: dobruskn@netvision.net.il

**Abstract**

The alternative approach to the displacement of gas/liquid equilibrium is developed on the basis of the Clapeyron equation. The phase transition in the system with well-established properties is taken as a reference process to search for the parameters of phase transition in the perturbed equilibrium system. The main equation, derived in the framework of both classical thermodynamics and statistical mechanics, establishes a correlation between variations of enthalpies of evaporation, $\Delta(\Delta H)$, which is induced by perturbations, and the equilibrium vapor pressures. The dissolution of a solute, changing the surface shape, and the effect of the external field of adsorbents are considered as the perturbing actions on the liquid phase. The model provides the unified method for studying (1) solutions, (2) membrane separations (3) surface phenomena, and (4) effect of the adsorption field; it leads to the useful relations between $\Delta(\Delta H)$, on the one hand, and the osmotic pressures, the Donnan potential, the surface curvature, and the pore structure, on the other hand. The value of $\Delta(\Delta H)$ has a clear physical meaning and gives a new insight into our understanding of the apparently different phenomena. The model is applicable if the change between entropies of the comparable gas phases is far more than the difference between entropies of the liquid phases.

## 1. Introduction

The present paper deals with the gas-liquid equilibrium and its displacement in response to the (1) formation of solutions, (2) membrane separation (3) changing the liquid shape, and (4) effect of the external field. Being the essential parts of the physical chemistry, these topics are discussed in any courses of chemical thermodynamics and physical chemistry;[1-8] but a new approach, which will be developed here, is different from the classical one. As a rule, in the framework of thermodynamics the displacement of equilibriums is considered on the basis of an assumed effect of physical and chemical disturbances on the chemical potentials of coexisting phases.[1-8] Our objective is to show that the theory of the displacement may be developed on the basis of the Clapeyron equation (CE), without resorting to the concept of chemical potential.

In the modern literature, it is generally accepted to derive the CE either from the Maxwell relations or from the Gibbs function and relevant chemical potentials. But it was not necessarily the case: Clapeyron came to his equation in 1834 when Maxwell was only tree-years-old and Gibbs had not been born yet. The CE was deduced in a straightforward way from the analysis of the Carnot cycle. As the equation is grounded in the fundamental first and second laws, the corollaries from the CE, as well as the CE itself, are based on the first principles. The review of approaches used in the derivation of the CE may be found in the papers of Wisniak[9] and Potter.[10]

The Clapeyron equation, $dp/dT=\Delta S/\Delta V$, is a thermodynamically exact equation that contains no approximations. Here, $p$ is the equilibrium vapor pressure, $T$ is the temperature of phase transition, $\Delta S=\Delta H/T$, $\Delta H$ and $\Delta V$ are the changes of entropy, enthalpy and volume, respectively, associated with a change of phase. In practical problems one may neglect the molar volume of the condensed phase relative to the molar volume of gaseous phase and approximates the latter by the ideal gas equation, *V=RT/p*, were *R* is the gas constant. Then the CE results in the Clausius-Clapeyron approximation:

$$\frac{dp}{p} = \frac{\Delta H}{RT^2} dT \qquad (1)$$

Taking the latent heat as constant over a sufficiently small temperature interval, the equation integrates giving the integral Clausius-Clapeyron equation

$$\ln p = const - \frac{\Delta H}{RT} \qquad (2)$$

which relates the temperature dependence of vapor pressure to the change of enthalpy of the phase transition. The Clausius-Clapeyron equation has been checked experimentally over a wide range of conditions in experiments on the vapor pressure of solids and liquids and in measurements of melting curves. All the experiments have shown it to be obeyed to a high order of accuracy. Its validity provides one of the most direct tests of the truth of the second law of thermodynamics.

Although it is a general practice to use the CE for a study of the temperature dependence of equilibrium parameters, a potential of the equation for an account of the effects of other physical and chemical parameters except temperature has escaped notice. We intend to make up for a deficiency and, proceeding from the CE, to develop the model which describes the displacement of the equilibrium parameters due to the formation of solutions, membrane separation, changing the liquid shape, and the effect of an external field. We believe that the relations derived from the Clapeyron equation provide new insight into the problems and can be generalized to other equilibrium phenomena. The remainder of the paper is organized as follows. In the next section, we present the thermodynamic derivation of the main equation which establishes the correlation between variations of enthalpy and the saturation pressure.

The effect of variations of enthalpy on the osmotic pressures and the Donnan equilibrium is discussed in section 3. The following two sections are devoted to the effects of surface curvature and external field on the enthalpy of evaporations (desorption); besides, the statistical mechanical derivation of the main equation will be presented in section 5.3. Experimental data on a capillary condensation and adsorption are compared against calculations in section 6. In the final two sections, the unified approach to perturbations of equilibrium of liquids is discussed and the conclusions are drawn.

**2. Enthalpy of Evaporation and Equilibrium Vapor Pressures. Main Equation.**

Our objective is to deduce the main equation that will be used for examining a host of phenomena. We shall introduce the equation for the case of solutions proceeding from the Clapeyron equation, but further the statistical mechanical derivation of the main equation will be given in section 5.3.

**2. 1 Solutions of nonvolatile compounds**. Being an immediate corollary of the fundamental first and second laws, the Clapeyron equation is valid for any equilibrium system and, in particular, for the following: (1) for the system which contains a solution of nonvolatile inorganic compound of interest in equilibrium with its vapor pressure (for instance, an aqueous solution of NaCl). The parameters of this system will be further supplied with the subscript *sol* and (2) for the system with the pure solvent (water) in equilibrium with vapor. The solvent may be thought of as a solution with zero concentration. The system with the pure solvent plays the role of a reference system. Parameters of the reference system will be marked by the subscript *ref*. In these cases, the CE equation takes the following forms:

$$\frac{dp}{p_{sol}} = \frac{\Delta H_{sol}}{R} \frac{dT}{T^2} \qquad (3)$$

and

$$\frac{dp}{p_{ref}} = \frac{\Delta H_{ref}}{R}\frac{dT}{T^2} \tag{4}$$

where $p_{sol}$ and $p_{ref}$ are the equilibrium pressures over the solution and over the solvent, respectively, $\Delta H_{ref}$ is enthalpy of the solvent evaporation from the pure liquid, and $\Delta H_{sol}$ is that from the solution. Note that $p_{ref}$ is just the solvent saturation pressure $p_s$. Subtracting eq 4 from eq 3 one obtains

$$d \ln \frac{p_{sol}}{p_s} = \frac{\Delta(\Delta H)}{R}\frac{dT}{T^2} \tag{5}$$

Here $\Delta(\Delta H) = \Delta H_{sol} - \Delta H_{ref}$ is the difference between enthalpies of evaporation. To integrate eq. 5, the dependence of enthalpies on temperature must be introduced. Although either enthalpies of evaporation, $\Delta H_{sol}$ or $\Delta H_{ref}$, vary with temperature, their difference $\Delta(\Delta H)$ is supposed to be practically independent of temperature, since the temperature changes, to a great extent, cancel out each other by the operation of subtraction. In any case, the assumption about the constancy of the difference, $\Delta(\Delta H)$, should be closer to the observation than that in respect to either term, $\Delta H_{sol}$ or $\Delta H_{ref}$, taken alone. Assuming $\Delta(\Delta H)$ to be constant, an integration of eq 5 leads to $\ln p_{sol}/p_s = -\Delta(\Delta H)/RT + C$, where C is the integration constant. The constant of integration here is equal to zero, because at any temperature $p_{sol}$ approaches $p_s$ when $\Delta(\Delta H)$ tends to zero. Finally, one obtains

$$RT \ln \frac{p}{p_s} = -\Delta(\Delta H) \tag{6}$$

Here the subscript *sol* to the symbol for pressure is omitted, since $p$ usually denotes just the equilibrium pressure over solutions. Equation 6 is the main equation which correlates the enthalpy of evaporation with the equilibrium vapor pressure. As $p_s$ is a function of a solution concentration, the equation enables one to study the effect of concentration of solutes on enthalpy of evaporation. It shows that the activity coefficient of a solvent, *a*, defined as $a=p/p_s$ is given by the following expression:

$$a = Exp\{-\frac{\Delta(\Delta H)}{RT}\} \qquad (7)$$

**2.2 Solutions of volatile substances.** Consider, for the simplicity, a binary solution. The same reasoning as in the previous section results in two relations, one for each component, which are similar to eq 6. For example, the equilibrium pressures over water-acetone solution are described by the following system of equations:

$$RT \ln \frac{p^w}{p_s^w} = -\Delta^w(\Delta H) \qquad (8a)$$

$$RT \ln \frac{p^{ac}}{p_s^{ac}} = -\Delta^{ac}(\Delta H) \qquad (8b)$$

where superscripts *w* and *ac* refer to water and acetone, respectively, $p^w$ and $p^{ac}$ are the partial vapor pressures; the values $\Delta^w(\Delta H)$ and $\Delta^{ac}(\Delta H)$ are the differences between enthalpies of evaporation of the components from the solution and from the corresponding pure liquid substances.

**2.3 Effect of concentration on variations of enthalpy of evaporation.** Consider application of eq 6 to the analysis of the literature experimental data. First of all, let us verify the assumption about the independency of $\Delta(\Delta H)$ of temperature. The tabulated experimental data for the verification in the form of equilibrium vapor pressures over aqueous solutions versus variable concentrations of solutes are taken from *International Critical Tables* and *Chemical Handbook*.[11, 12] The values of $\Delta(\Delta H)$ are computed by eq 6; concentrations of solutes are expressed in molality, *m* (mol/kg of water). Figures 1 and 2 demonstrate that $\Delta(\Delta H)$ is really independent of temperature in a wide concentration range. The same conclusion is drawn for many other aqueous solutions ($H_2SO_4$, KOH, NaOH, etc.). For some neutral solutions of strong electrolytes, (NaCl and NaNO$_3$) one can observe a small reduction of $\Delta(\Delta H)$ with temperature; however, even in these cases $\Delta(\Delta H)$ remains practically constant in a narrower interval. Hence, the main assumption is corroborated by the experimental data.

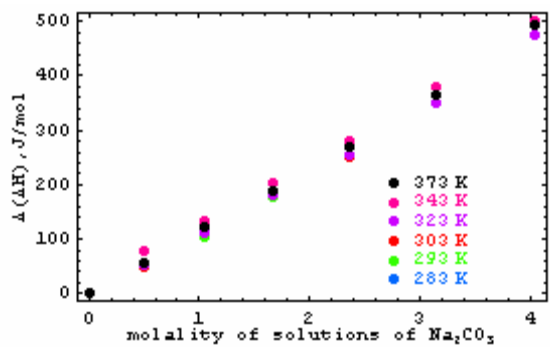

**Figure 1.** Effect of temperature on the variation of enthalpies of water evaporation $\Delta(\Delta H)$ from solutions of sodium carbonate.

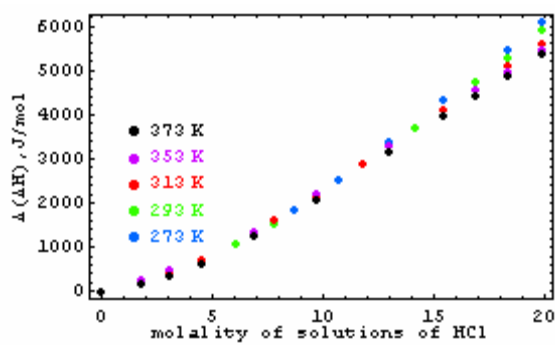

**Figure 2.** Effect of temperature on the variation of enthalpies of water evaporation $\Delta(\Delta H)$ from solutions of hydrochloric acid.

The changes of enthalpies of evaporation of aqueous solutions at 373.15 K with molality of chlorides of metals of the first and second groups of the periodic table are demonstrated in Figures 3 and 4. The absolute values of enthalpies of water evaporation ($\Delta H_{abs}$) from solutions at 373.15 K are equal to $\Delta H_{abs} = 40700 + \Delta(\Delta H)$ J/mol, where 40700 J/mol is the enthalpy of water evaporation at 373.15 K.

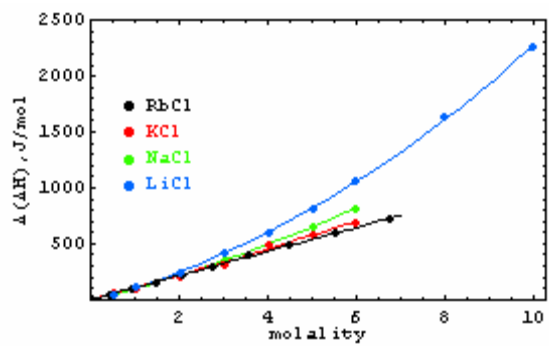

**Figure 3**. Effect of concentration of chlorides of the first group metals on the variation of enthalpies of water evaporation Δ(ΔH) at 373.15 K.

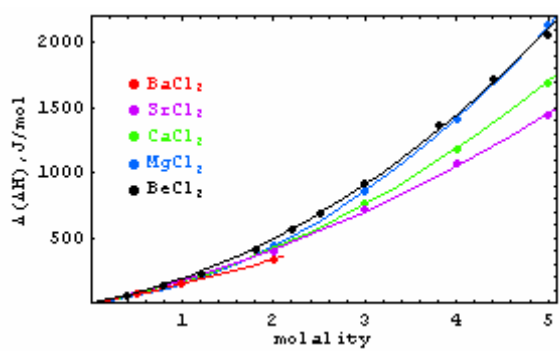

**Figure 4.** Effect of concentration of chlorides of the second group metals on the variation of enthalpies of water evaporation Δ($\Delta H$) at 373.15 K.

One may see that the energy expenditure for evaporation of one mole of water increases with the concentration of solution and, hence, dissolved inorganic compounds enhance the water/solution interactions and the retaining capacity of liquid phase. The water/solution interactions are enhanced with decreasing cation radii in the following sequences: Li>Na>K>Rb and Be≥Mg>Ca>Sr>Ba.

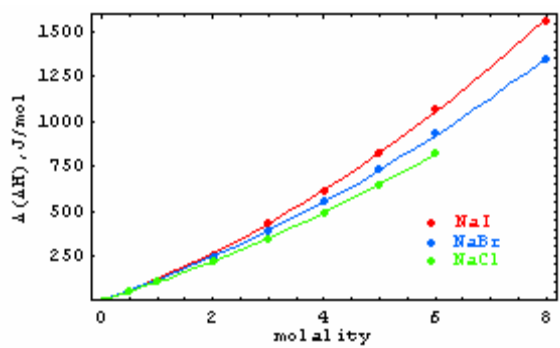

**Figure 5.** Effect of concentration of sodium halogenides on the variation of enthalpies of water evaporation Δ(Δ*H*) at 373.15 K.

The effect of anion radius is to some extent opposite: keeping the cation radius constant (Figure 5), the interactions and, hence, the retaining capacity enhances with increasing anion radius. At first glance it seems that there may be a correlation between the Δ(Δ*H*) and enthalpies of ion hydrotation, Δ$H_{hyd}$. However, enthalpies of ion hydrotation and enthalpies of water evaporation from ionic solutions relate to quite different interactions: the former characterizes the interactions of the definite ions with water, whereas the latter describes the interactions of water molecules with all species in the solution (ions, water molecules, products of hydrolysis). The magnitudes of these values are entirely different and Δ$H_{hyd}$ is hundreds times grater than Δ(Δ*H*).

Variations of Δ(Δ*H*) for organic ideal and non-ideal binary solutions are plotted in Figures 6 and 7. Experimental data are taken from the literature.[11, 13]

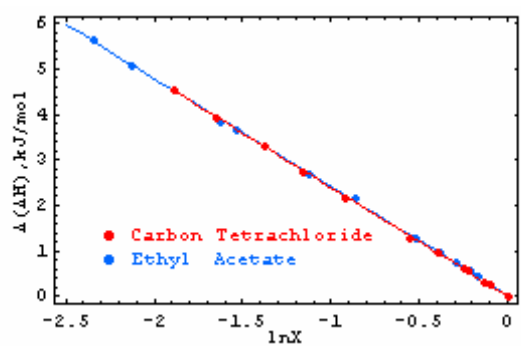

**Figure 6.** Variations in enthalpy of evaporation Δ(Δ*H*) of components at 323.14 K in the ideal mixture of Carbon Tetrachloride and Ethyl Acetate versus logarithm of mole fractions ln*X*. The red line and points describe Δ(Δ*H*) of $CCl_4$ against the mole fractions of $CCl_4$, whereas Δ(Δ*H*) of Ethyl Acetate (blue line and blue points) is given versus the mole fractions of Ethyl Acetate.

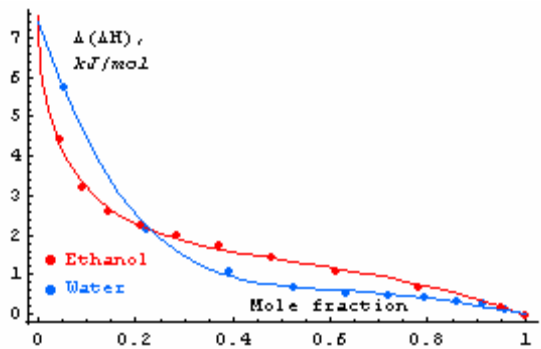

**Figure 7.** Variations of enthalpy of evaporation of components Δ(Δ$H$) versus mole fractions in the non-ideal mixture of Ethanol and Water at 348.15 K.

One may see that: (1) variations of enthalpies of each components increase with the growth of concentration of the second component; (2) for the ideal solution the plots related to the first and second components practically coincide with one another; and (3) for the non-ideal solution, the cross-points and flex-points appear on the curves and values of Δ(Δ$H$) of the individual components of solutions do not coincide with one another. Hence, for the ideal solution the resistance to evaporation for each component is equal to one another, while for non-ideal solutions there are different resistances to evaporation for each component of the solution.

**3. Membrane Equilibrium**

**3.1 Relation between osmotic pressure and enthalpy of evaporation.** Consider two vessels within a box at constant temperature, one containing water and another containing an aqueous solution of nonvolatile substance (Figure 8). The vapor phase in the box acts as a semi-permeable membrane which allows water vapor to flow freely between the vessels but blocks the movement of the solute. As the saturation water pressure of pure water $p_s$ exceeds that of a solution, water vapor condenses in the left tube resulting in an increase of the level of liquid and the hydrostatic pressure which the liquid column exerts on the solutions.

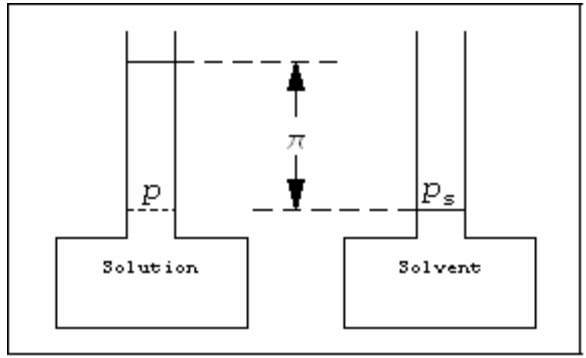

**Figure 8.** Two open vessels, one containing only the solvent (water) and another containing an aqueous solution of nonvolatile substance, placed in the closed box.

We suppose that volumes of vessels are so large that that dilution of the solution is negligible. When static pressure is built up within a fluid, the Pascal law postulates that it will be distributed equally within the fluid volume and perpendicular to the surface of the fluid boundary. The question we ask is: how much must the external pressure change to keep the solution in equilibrium with the pure solvent?

To answer the question, we should take into account the following: the difference in saturation pressures originates from the difference in enthalpies of evaporation $\Delta(\Delta H)$ which, in turn, results from the enhanced retaining capacity of solutions due to the water-ions interactions. However, enthalpy is influenced not only by the presence of a solute; it, for instance, can be changed by the hydrostatic pressure exerted on the solution. Quantitatively, the effect of pressure on enthalpy is found from thermodynamics. Keeping in mind one of the Maxwell relations,

$$\left(\frac{\partial S}{\partial p}\right)_T = -\left(\frac{\partial V}{\partial T}\right)_P \tag{9}$$

differentiation of the equation

$$dH = TdS + Vdp \tag{10}$$

in respect $p$ at constant $T$ leads to the following expression

$$\left(\frac{\partial H}{\partial p}\right)_T = V - T\left(\frac{\partial V}{\partial T}\right)_p \tag{11}$$

which describes the effect of pressure on enthalpy.[14] Using the constant molar volume assumption ($(\partial V/\partial T)_p = 0$), one obtains the following equation for the effect of pressure on the enthalpy of evaporation:

$$\Delta H = \int_{p_1}^{p_2} V dp \tag{12}$$

Assuming that volume of liquids is also independent of pressure, we have

$$\Delta H = (p_2 - p_1)V = V\Delta p \tag{13}$$

Hence, condensation in the left tube continues until the growing hydrostatic pressure, which is generated by the difference in the water levels, results in an increase in enthalpy that counterbalances its lowering due to the effect of solute. Denoting $\Delta p$ by the symbol $\pi$, we have from eq 13

$$\pi V = \Delta(\Delta H) \tag{14}$$

where $\Delta(\Delta H)$ is just a decrease of enthalpy due to interactions with a solute. From here the osmotic pressure, $\pi$, is equal to

$$\pi = \frac{\Delta(\Delta H)}{V} \tag{15}$$

Substituting $\Delta(\Delta H)$ from eq 6 into eq 15 gives another expression for the osmotic pressure

$$\pi = \frac{RT}{V}\ln\frac{p_s}{p} \tag{16}$$

which relates the osmotic pressure with the equilibrium pressure over the solution established in the absence of the membrane. Pay attention to that eq 16 by no means means that the solvent exerts an osmotic pressure on the solution and simultaneously the reduced equilibrium pressure is established over the solution. It just means that lowering enthalpy on dissolving leads either to the equilibrium pressure depletion or to the osmotic pressure on the solution, depending on the system arrangement. The derivation of eq. 15 shows that osmotic pressure is determined by the properties of solvent; hence, *V* is its partial molar volume. As the position of equilibrium is independent of the path towards it, the same result is held for the connected vessels separated by the real semi-permeable membrane, and not by the "semi-permeable" gas phase.

Compare the experimental osmotic pressures in solutions with those computed by eq 16 (Figure 9). For this purpose, the vast experimental data of Robinson and Stokes on solutions of electrolytes is used.[15] Note that the authors, instead of reporting the experimental values of osmotic pressures, give the osmotic coefficients, φ, bound to the osmotic pressure of aqueous solutions by the relation

$$\pi = \frac{RT \nu m M \varphi}{1000 V} \qquad (17)$$

where ν is a number of moles of ions formed from 1 mole of electrolyte, *M* is the molecular mass of water, *m* is the molality, and *V* is the partial molar volume of water. The experimental values of osmotic pressures are recalculated from the reported values of φ by eq 17. The authors use the value of $V=18.01$ cm$^3$/ gram-mole.

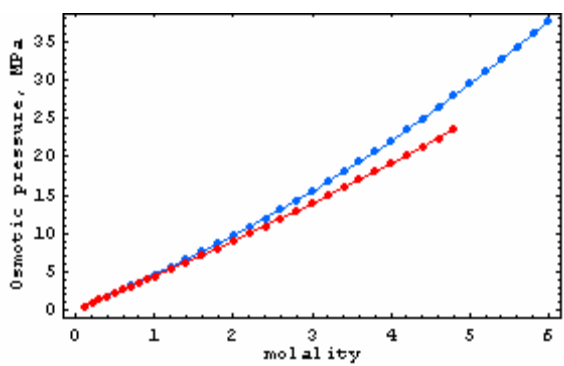

Figure 9. Experimental (lines) and calculated by eq 16 (points) osmotic pressures of solutions of NaCl (blue) and KCl (red) at 298 K.

A comparison of these experimental osmotic pressures with those calculated by eq 16 from the experimental values of the relative pressures (Figure 9) shows that there is a very good agreement between both series of values.

As measurements of osmotic pressures are more accurate than measurements of vapor pressures, these former will be further used for computation of $\Delta(\Delta H)$. Variations of $\Delta(\Delta H)$ of water with changing concentrations of electrolytes in the range 0 to 20 mol/kg are shown in Figure 10.

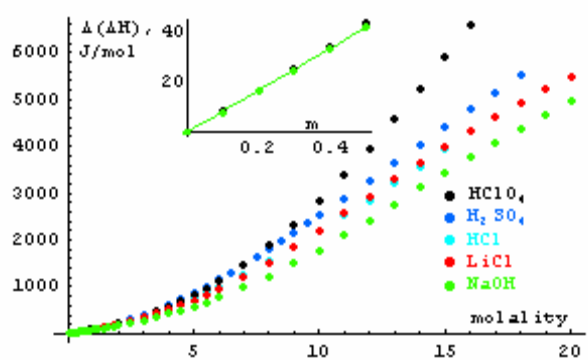

**Figure 10.** Effect of solute concentrations on variations of enthalpy of evaporation $\Delta(\Delta H)$ at 298.15 K. Values of $\Delta(\Delta H)$ for solutions of HCl and LiCl practically coincide.

It is seen that for dilute solutions changing enthalpy at constant temperature is directly proportional to the solute concentration and independent of its chemical identity. Hence, $\Delta(\Delta H)$ is one of the colligative properties of solutions. For $m>0.1$-$0.5$, $\Delta(\Delta H)$ does depend on the chemical composition and changes in the sequence: $NaClO_4$ >$H_2SO_4$ >HCl ≈LiCl >NaOH. In conclusion of this section, we report the values of

Δ(ΔH) for water at 298.15 K for 85 electrolytes (Tables 1-3) calculated on the basis of experimental data of Robinson and Stokes.

**Table 1.** Variations of enthalpy of water evaporation for 1:1 electrolytes with molality m= 0.1 mol/kg and m=3 mol/kg

| Solute | Δ(ΔH), J/mol | | Solute | Δ(ΔH), J/mol | |
|---|---|---|---|---|---|
| | $m=0.1$ | $m=3$ | | $m=0.1$ | $m=3$ |
| HCl | 8.40 | 360 | Na Hsuc | 8.23 | 242 |
| HBr | 8.44 | 394 | Na HAd | 8.29 | - |
| HI | 8.48 | 410 | NaTol | 8.23 | 200 |
| HClO$_4$ | 8.43 | 471 | NaCNS | 8.34 | 290 |
| HNO$_3$ | 8.37 | 308 | NaH$_2$PO$_4$ | 8.11 | 186 |
| LiOH | 7.96 | 236 | KOH | 8.31 | 333 |
| LiCl | 8.36 | 343 | KF | 8.28 | 280 |
| LiBr | 8.39 | 364 | KCl | 8.25 | 250 |
| LiI | 8.48 | 392 | KBr | 8.26 | 255 |
| LiClO$_4$ | 8.47 | 379 | KI | 8.30 | 264 |
| LiNO$_3$ | 8.35 | 315 | KClO$_3$ | 8.13 | - |
| LiAc | 8.32 | 292 | KBrO$_3$ | 8.10 | - |
| Li Tol | 8.26 | 244 | KNO$_3$ | 8.07 | 161 |
| NaOH | 8.24 | 292 | K Ac | 8.40 | 328 |
| NaF | 8.23 | - | K HMal | 8.19 | 210 |
| NaCl | 8.30 | 279 | K HSuc | 8.21 | 338 |
| NaBr | 8.32 | 296 | K HAd | 8.26 | - |
| NaI | 8.35 | 317 | K Tol | 8.20 | 170 |
| NaClO$_3$ | 8.25 | 235 | NH$_4$Cl | 8.25 | 247 |
| NaClO$_4$ | 8.28 | 256 | NH$_4$NO$_3$ | 8.11 | 198 |
| NaBrO$_3$ | 8.17 | - | RbCl | 8.22 | 245 |

| NaNO$_3$ | 8.20 | 216 | RbNO$_3$ | 8.04 | 157 |
| Na For | 8.29 | 268 | CsCl | 8.16 | 235 |
| Na Ac | 8.37 | 315 | CsNO$_3$ | 8.03 | - |
| NaHmal | 8.22 | 228 | | | |

Ac=acetate; Ad=adipate; Suc=succinate; For=formate; Mal=malonate; Tol=p-toluenesulphonate;

To the first approximation, a decrease of enthalpy of evaporation at concentration of 0.1 mol/kg for 1,1 electrolytes may be considered to be independent of the chemical identity and equal to

$$\Delta(\Delta H) \approx 83m \qquad (18)$$

The chemical identity begins to make its appearance at higher concentrations. In the case of $m$=3 mol/kg, the effect of ions is summarized in Table 2. It shows how counterions of a given ion affect the absolute values of enthalpies of water evaporation.

**Table 2.** Effect of inorganic counterions on interactions in aqueous solutions at $m$=3 mol/kg

| Ion | Effect of counterions | Ion | Effect of counterions |
|---|---|---|---|
| H$^+$ | ClO$_4^-$>I$^-$>Br$^-$>Cl$^-$>NO$_3^-$ | OH$^-$ | K$^+$>Na$^+$>Li$^+$ |
| Li$^+$ | I$^-$> ClO$_4^-$>Br$^-$>Cl$^-$> NO$_3^-$>OH$^-$ | Cl$^-$ | H$^+$>Li$^+$>Na$^+$>K$^+$>Rb$^+$>Cs$^+$ |
| Na$^+$ | I$^-$>OH$^-$>Br$^-$>Cl$^-$> ClO$_4^-$> ClO$_3^-$> NO$_3^-$ | Br$^-$ | H$^+$>Li$^+$>Na$^+$>K$^+$ |
| K$^+$ | OH$^-$>F$^-$>I$^-$> Br$^-$>Cl$^-$> NO$_3^-$ | I$^-$ | H$^+$>Li$^+$>Na$^+$>K$^+$ |

In the case of polyvalent electrolytes, the values of $\Delta(\Delta H)$ are close to each other at the constant anyone and $m$=0.1 mol/kg, irrespective of cations ($\approx$11.4 for MCl$_2$, 14.0-14.5 for MCl$_3$, and so on); for high concentration of polyvalent electrolytes, the experimental data are scarce (Table 3).

**Table 3.** Variation of enthalpy of water evaporation for polyvalent electrolytes with molality $m=0,1$ and $m=3$.

| Solute | Δ(ΔH), J/mol | | Solute | Δ(ΔH), J/mol | |
|---|---|---|---|---|---|
| | $m=0.1$ | $m=3$ | | $m=0.1$ | $m=3$ |
| $MgCl_2$ | 11.50 | 805 | $K_2SO_4$ | 10.40 | - |
| $MgNO_3$ | 11.45 | 685 | $(NH_4)_2SO_4$ | 10.24 | 254 |
| $CaCl_2$ | 11.40 | 713 | $AlCl_3$ | 14.59 | - |
| $SrCl_2$ | 11.35 | 653 | $ScCl_3$ | 14.19 | - |
| $BaCl_2$ | 11.26 | - | $CrCl_3$ | 14.19 | - |
| $MnCl_2$ | 11.39 | 583 | $LaCl_3$ | 14.03 | - |
| $FeCl_2$ | 11.40 | - | $SmCl_3$ | 14.05 | - |
| $CoCl_2$ | 11.44 | 685 | $K_3[Fe(CN)_6]$ | 12.95 | - |
| $CuCl_2$ | 11.28 | 453 | $K_4[Fe(CN)_6]$ | 13.24 | - |
| $ZnCl_2$ | 11.31 | 344 | $Al_2(SO_4)_3$ | 9.35 | - |
| $Li_2SO_4$ | 10.92 | 394 | $Cr_2(SO_4)_3$ | 9.21 | - |
| $Na_2SO_4$ | 10.59 | 325 | $Tc(NO_3)_4$ | 15.02 | 1032 |
| $Na_2CrO_4$ | 10.87 | 362 | | | |

**3.2 Relation between enthalpy of evaporation and the Donnan potential.** In 1911, Donnan[16] proposed a theory of equilibrium between solutions separated by a specific membrane, which is permeable for the solvent and for one kind of ions and non-permeable for the counterions. In such a system the electric potential difference is generally established between the membrane and the solutions. Consider, for example, an aqueous solution of salt NaR (Congo red in Donnan' study) with concentration $n_{os}$ in contact with a semi-permeable membrane which is impermeable to the anion R⁻ and separates the Congo red solution in compartment I from water in compartment II. The diffusion of Na$^+$ from compartment I to the compartment II generates a difference of electric potential across the membrane. The membrane surface bears negative charges R$^-$ in the solution of NaR and positive charge of Na$^+$ in the water compartment. The potential difference attains equilibrium when the electrical attraction of solution ceases the diffusion of Na+ down its concentration gradient. The correlation between the Donnan potential and changing of enthalpy of evaporation may be found as follows.

The membrane can be thought of as a parallel-plate capacitor placed in the solution. The two charges of the capacitor walls, +Q and –Q, cause the plates to attract each other exerting pressure on the solution. The force of the electric attraction is $f = Q^2/2a\varepsilon$, where $a$ is the area of the membrane and $\varepsilon$ is the dielectric permittivity of water. Hence, the additional pressure, $p=f/a$, is equal to[17, 18]

$$p = \frac{q_s^2}{2\varepsilon} \qquad (19)$$

where $q_s=(Q/a)$ is the surface charge density. The solute induces the reduction of enthalpy $\Delta(\Delta H)$ which, in the case of ion non-permeable membrane, determines the osmotic pressure $\pi = \Delta(\Delta H)/V_{mol}$. In our case of an ion semi-permeable membrane, the same effect is brought about by the "electrostatic" pressure $p$ exerted by the membrane on the solution: the system will be in equilibrium when $p=\pi$, that is, $q_s^2/2\varepsilon = \Delta(\Delta H)/V_{mol}$. From here, the equilibrium charge density on the membrane is equal to

$$q_s = \sqrt{\frac{2\varepsilon \times \Delta(\Delta H)}{V_{mol}}} \qquad (20)$$

So far we have discussed the equilibrium between pure water and the solution which contains one kind of non-permeable ions with concentration $n_{os}$. Now suppose that in the foreign electrolyte is added in the system which does penetrate through the semi-permeable membrane and its concentration is $n_p$. Since its concentrations in both compartments at equilibrium are equal to one another, (1) the foreign electrolyte does not generate the difference of osmotic pressures between the compartments, (2) the value of $\pi$ and $\Delta(\Delta H)$ will continue to be determined by $n_{os}$, and (3) eq 20 remains valid. Nevertheless, the foreign electrolyte affects the distribution of the potential in a solution. This problem has been intensively studied for decades. When a membrane that is permeable to electrolyte ions and contains charged groups at uniform density $q_s$ is in equilibrium with a symmetrical electrolyte solution of concentration $n$ and valence Z, the Donnan potential $\varphi$ relative to the bulk solution is given by the following expression[19-21]

$$\varphi = \frac{kT}{Ze}\ln\left\{\frac{q_s}{2Zn} + \left[\left(\frac{q_s}{2Zn}\right)^2 + 1\right]^{1/2}\right\} = \frac{kT}{Ze}\text{arcsinh}\left(\frac{q_s}{2Zn}\right) \quad (21)$$

where $k$ is the Boltzmann constant and $e$ is the elementary electric charge. Substitution of the expression for $q_s$ gives

$$\varphi = \frac{kT}{Ze}\text{arcsinh}\left\{\frac{[2\varepsilon \times \Delta(\Delta H)]^{1/2}}{2ZnV_{mol}^{1/2}}\right\} \quad (22)$$

Note that $n$ is the sum of $n_{os}$ and $n_p$.

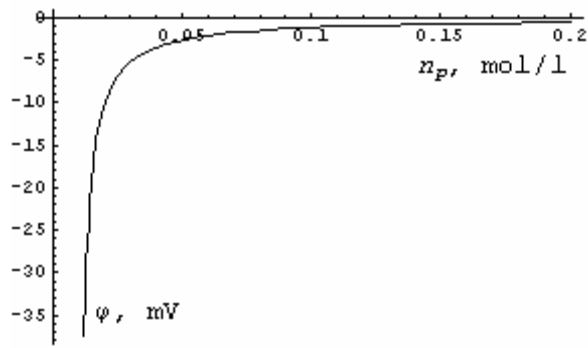

**Figure 11.** Effect of concentration of 1:1 foreign electrolytes $n_p$ on the Donnan

potential φ in aqueous solution at *T*=298.15 K. The concentration of a nonpermeating ion with the charge of (-1) is equal to $n_{os}$=0.01 mol/l. When $n_p$=0, φ=-55 mV.

Figure 11 shows that the Donnan potential sharply decreases with increasing $n_p$. For small concentration of $n_{os}$, the value of Δ(Δ*H*) may be approximately calculated as Δ(Δ*H*)≈83*m* (eq 18).

## 4. Effect of Surface Curvature on Δ(Δ*H*) and Equilibrium Pressures

The equilibrium vapor pressures over a curved liquid surface are described by the same equation 6, which is derived in similar way as in the case of solutions: it is enough to substitute the word "solution" by the words "sample with a curved surface." In this case, Δ(Δ*H*) is the difference of enthalpies of evaporation from the bulk liquid and from the sample, and computation of the equilibrium pressure is reduced to the calculation of Δ(Δ*H*). In contrast to the solutions for which there is no a reliable molecular model for calculating Δ(Δ*H*), such a model has been developed in the cases of liquid samples with curved surfaces and the predicted equilibrium pressures can be compared with observations. In further text, we will use the word droplet as a shorthand for any liquid sample with a curved surface.

**Effect of curvature on Δ(Δ*H*).** Our objective is to find the difference between enthalpy of evaporation from a bulk liquid and enthalpy of evaporation from a droplet of the same chemical nature. Consider a simple molecular model of evaporation, which takes into account the surface phenomena and can shed light on their effect on variations of enthalpy with the curvature.[22-24] Since Δ(Δ*H*) is sought at the same temperatures of the liquids samples, the kinetic energies of species in both samples, $E^s_{kin}$, are equal to one another and only the changes of the potential energy, Δ$E_{pot}$, are significant. Keeping in mind that for condensed phases the changes of internal energy, Δ*U*, and Δ*H* are practically identical, one obtains the sequence of equalities:

$$\Delta(\Delta H) \cong \Delta(\Delta U) = \Delta E_{pot} \qquad (23)$$

Since changing the energies on condensation or evaporation are equal in value but opposite in sign, we, for the sake of convenience, shall continue the discussion in terms of internal energy of condensation.

Because $U$ is the state function, one can choose any convenient way between initial and final states for calculating $\Delta U$. In particular, imagine that condensation consists of three consecutive stages: in the first stage, each gas molecule is adsorbed on the surface and then, in the second and third stages, it penetrates the surface layer and, finally, into the interior of the liquid. The picture that emerges from the model is that of a quiescent liquid surface, while it is actually in the state of violent agitation on the molecular scale with individual molecules passing back and forth between the surface and the bulk regions on either side. As Adamson[25] writes, "under a microscope of suitable magnification, the surface region should appear as a fuzzy blur, with the average density varying in some continuous manner from that of the bulk phase to that of the vapor phase." It may appear that there is a conflict between our model and the reality. In this connection, the following should be taken into account: (1) From the relation $\Delta(\Delta H) \approx \Delta E_{pot}$ (eq 23) follows that the changes of enthalpy may be described in terms of motionless molecules; (2) The model does not pretend to be the model of surface region. It just takes advantage of the path-independence of enthalpy and introduces the hypothetical intermediate states with surface tensions of real objects.

The first stage of the model is autoadsorption and its energy effect is the energy of autoadsorption. This term denotes adsorption of vapor on the surface of its own condensed phase (for example, water vapors on the surfaces of either ice or liquid water) when molecules only touch the surface without entering the surface layer. In the parlance of adsorption theory, [26, 27] the autoadsorption energy ε* is the energy of adsorption in the Henry limit (that is, at the limit of zero coverage); after the adsorption layer has been completed, each of the adsorbed molecules interacts not only with the species located beneath the layer but also with its lateral neighboring molecules.[26, 27] The energy of lateral interactions, ε*$_{lat}$, contributes to the total effect and determines a change of internal energy corresponding to the penetration in the surface layer. The energetic effect of the third stage when a molecule moves from the surface layer to the bulk liquid is equal in magnitude but opposite in sign to the total (excess) surface energy, $E_s$. The total surface energy is just defined as the energy gained by a molecule while being transferred from the bulk liquid to its surface;[7, 25, 28, 29] it is the excess energy of the molecules in the surface layer in respect to their energy in the volume. This value is in a close association with the free surface energy (the surface tension), γ:

$$E_s = \gamma - T \frac{d\gamma}{dT} \qquad (24)$$

where (-dγ/dT) is the excess surface entropy and T(-dγ/dT) is the quantity of latent heat absorbed in the reversible isothermal change of the surface area.[30] $E_s$ is nearly temperature independent in the wide range not too close to the critical temperature, $T_c$, but eventually drops to zero at $T_c$.[25]

The interplay between the change of internal energy on condensation, $\Delta U_{con}$, on one hand, and ε*, ε$_{lat}$, and $E_s$, on the other hand, is schematically depicted in Figure 12.

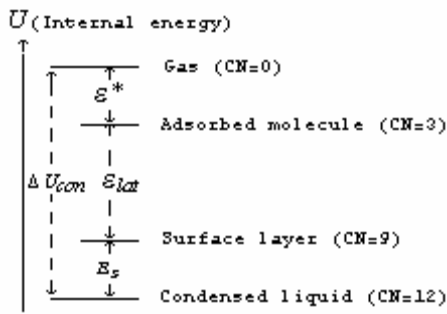

**Figure 12.** Schematic diagram of the energetic levels of molecules. $\Delta U_{con}$, $\varepsilon^*$, $\varepsilon_{lat}$, and $E_s$, are the energy of condensation, the autoadsorption energy, the energy of lateral interactions, and the excess surface energy, respectively. *CN* is the coordination number of a molecule in the case of a hypothetical closed packed liquid.

It is evident from Figure 12 that

$$\Delta U_{con} = -(\varepsilon^* + E_s + \varepsilon_{lat}^*) \qquad (25)$$

By convention, $\varepsilon^*$ is positive, whereas $\Delta U$ on condensation is negative. As accepted in the theory of adsorption, the asterisk $^*$ indicates that the parameter corresponds to the well depth. Then, for evaporation one obtains

$$\Delta U_{ev} = \varepsilon^* + E_s + \varepsilon_{lat}^* \qquad (26)$$

Applying equation 26 to the bulk liquid and to the liquid sample with the curved surface, one has

$$\Delta(\Delta H) \cong \Delta(\Delta U) = \Delta\varepsilon^* + \Delta E_s + \Delta\varepsilon_{lat}^* \qquad (27)$$

Hence, the total variation of enthalpy of evaporation is the sum of the change of autoadsorption energy $\Delta\varepsilon^*$, the variation of energy of lateral interactions $\Delta\varepsilon_{lat}$, and that of the excess surface energy $\Delta E_s$.

As an example, consider a model of a closed packed liquid. In the bulk and in the surface layer of such a liquid each of the molecules is surrounded by 12 and 9 nearest neighbors, respectively. A molecule adsorbed on the surface of close-packed spheres at the position of the minimum potential energy has only three nearest neighbors. One may see that the number of nearest neighbors lost by a molecule while moving from volume to surface (12-9=3) is equal to that on desorption into a gas phase (3-0=3). As the energetic properties are mainly determined by the interactions with the nearest neighbors, one might expect that the autoadsorption energy and the excess surface energy are equal to each other in magnitude:

$$E_s = \varepsilon^* \qquad (28)$$

These argumentations were put forward by Stefan[7] and Skapski;[31] the general evidence for this relation is given in our previous publication.[24, 32]

Let us now turn to the surface layers of droplets and their parent liquids. It is known that for a characteristic size of droplets more than 1 nm the droplet surface tension is close to that of bulk liquids.[25] It is possible only if the arrangements of their surface layers are identical. For such surface layers, the energetic effects of the second and third stages must be also identical and independent of radii. Hence, $\Delta\varepsilon_{lat}$, and $\Delta E_s$ in eq 27 are equal to zero and variations of enthalpy with radii are determined by the difference of the autoadsorption energies on the surfaces of the comparable samples

$$\Delta(\Delta H) = \Delta\varepsilon^* \qquad (29)$$

It is evident from Figure 13 that the energy autoadsorption $\varepsilon^*$ does depend on the curvature and $\Delta\varepsilon^* \neq 0$.

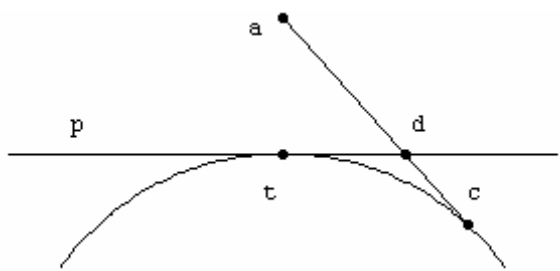

**Figure 13.** Schematic comparison of the autoadsorption energies. When a molecule located in a point *a* is adsorbed either on a planar surface represented by the line *p* or on a convex droplet surface depicted by the circular arc the separations *ac* between adsorbate *a* and points of a convex surface *c* are greater than the distances *ad* from the planar surface *p*, except for the tangential point *t*; as a consequence, the energy of adsorption on the planar surface exceeds that on the convex droplet surface.

In the general case, ε* can be the sum of energies of hydrogen bonding, $\varepsilon_{hb}$, dipole-dipole ($\varepsilon_{dd}$) and the London dispersion ($\varepsilon_{Ld}$) interactions, depending upon the chemical nature of liquids. From here, one has

$$\Delta \varepsilon^* = \Delta \varepsilon_{hb} + \Delta \varepsilon_{dd} + \Delta \varepsilon_{Ld} \tag{30}$$

It is clear that the hydrogen bond is the local chemical bond; it does not depend on the curvature and $\Delta\varepsilon_{hb}=0$. The dipole moments of molecules also do not depend on the liquid surface curvature; dipole-dipole interactions are non-additive and, as the arrangements of surface layers retain invariant, for the comparable surfaces $\Delta\varepsilon_{dd} \approx 0$. On the contrary, the London interactions are additive ones; they depend on the sample shape and dimension. Therefore, the changing adsorption energy $\Delta\varepsilon^*$ with the curvature is determined by the difference of the energies of dispersion interactions between (1) the vapor molecules and the bulk liquid and (2) the vapor molecules and the droplet:

$$\Delta \varepsilon^* = \Delta \varepsilon_{Ld} \tag{31}$$

The equation itself provides the powerful approximation method for calculating $\Delta\varepsilon^*$ and makes the problem tractable. For example, even in the case of water for which the main contributions to the autoadsorption energies come from the hydrogen bonds and dipole-dipole interactions, $\Delta\varepsilon^*$ is determined only the difference between the energies of dispersion interactions of water molecule with the flat water surface and with the curved water surface, respectively. The theory of the dispersion interactions is well developed and one can use the Lennard-Jones (LJ) potential for calculating of Δε*.

The expressions for calculating Δε* were derived in our previous publications and we report here only the final results. The physical reasoning has shown that the required values are given as follows

$$\Delta\varepsilon^* \equiv \Delta\varepsilon_{LJ} = E_s - E_s \varepsilon^*_{sam}/\varepsilon^*_{slab} \qquad (32)$$

where $\varepsilon^*_{sam}$ is the LJ-energy of interactions between the vapor molecule and the liquid sample of interest and $\varepsilon^*_{slab}$ is the LJ-energy of interactions between the vapor molecule and the bulk liquid which is modeled by a semi-infinite slab.[22-24] The ratio of the dispersion components (a dispersion ratio), $\varepsilon^*_{sam}/\varepsilon^*_{slab}$, for a sphere and a semi-infinite slab, $\varepsilon^*_{sp}/\varepsilon^*_{slab}$, is known from the adsorption theory:[22-24]

$$\frac{\varepsilon_{sp}(r,z)}{\varepsilon^*_{slab}} = \frac{24r^3}{\sqrt{10}} \left\{ \frac{1}{\left((r+z)^2 - r^2\right)^3} - \frac{15(r+z)^6 + 63(r+z)^4 r^2 + 45(r+z)^2 r^4 + 5r^6}{15\left[(r+z)^2 - r^2\right]^9} \right\}$$

(33)

where $r$ is the reduced radius of the sphere and $z$ is the reduced distance of the molecule from the sphere. Note, that all sizes are expressed here in the reduced forms with the Lennard-Jones diameter, σ, as a scale parameter (for example, if $R$ is the absolute radius of a droplet, then $r=R/\sigma$). For a given $r$, the extremum, $\varepsilon^*_{sp}(r, z^*)$, is found by the numerical method with respect to z; it usually occurs at $z^* \approx 0.858$. In the case of adsorption on the internal surface of the infinite cylindrical liquid capillary with a reduced radius $r$, the extremum value of the ratio is given as follows:

$$\frac{\varepsilon^*_{cyl}}{\varepsilon^*_{slab}} = \frac{27}{2\sqrt{10}}\pi\left\{\frac{21}{288r^9}\,_2F_1\left[\frac{9}{2},\frac{11}{2};1;\frac{(r-0.858)^2}{r^2}\right] - \frac{1}{3r^3}\,_2F_1\left[\frac{3}{2},\frac{5}{2};1;\frac{(r-0.858)^2}{r^2}\right]\right\}$$

(34)

were $_2F_1$ is the hypergeometric function. For liquid samples in the form of figures of revolution, the procedure of calculating $\varepsilon^*_{sam}/\varepsilon^*_{slab}$ is given in ref 33. Such nonspherical liquid islands are formed in the case of adsorption on porous adsorbents. The application of the model for calculating $\Delta(\Delta H)$, together with eq 6, ensures the computation of equilibrium pressures over the curved surfaces.

**5. Effect of Adsorption Field on Equilibrium Pressure of Adsorbed Liquid.**

Our objective is to show that the main equation describes adsorption on porous adsorbents. To do this, let us first examine the characteristic feature of adsorption in porous media and show that it may be considered as an evolution of two-dimensional condensation (2DC) that occurs on the surface of pore walls.

**5.1 Adsorption on a flat surface**. The important molecular principles underlying adsorption phenomenon have been cleared up by the statistical mechanical theories of adsorption, which establishes correlations between observed properties of adsorption films and energies of molecular interactions.[26, 27, 34, 35] It has been shown that, irrespective of models of adsorption, the attractive adsorbate-adsorbate (lateral) interactions must be involved to predict an adsorption behavior. On increasing of energies of lateral interactions, the isotherm curve passes through a region where it may be approximated by a straight line normal to the $p$-axis (Figure 14). When not only the nearest-neighbor, but also the next-nearest-neighbor interactions are taken into account,[36, 37] the fractional adsorption, $\theta$, increases rapidly with the pressure and the cooperative effect sets at a low surface coverage and terminates at a higher surface coverage in the range

$$0.05\text{-}0.1 < \theta < 0.9\text{-}0.95 \tag{35}$$

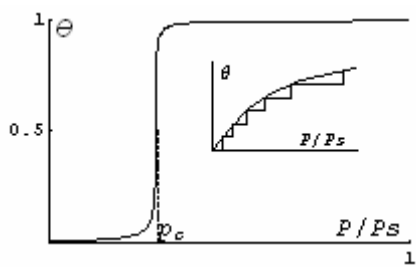

**Figure 14**. Schematic isotherm of adsorption on a homogeneous surface. θ is the fractional adsorption, $p_c$ is the critical pressure of the 2DC, and $p/p_s$ is the relative pressure. **Insert**. Adsorption on a heterogeneous sample may be presented as the sum of adsorption on the individual homogeneous areas.

The above straight line (Figures 14) represents a region where a two-dimensional condensation (2DC) (that is, a phase transition) occurs at the critical condensation pressure, $p_c$. Since for bulk adsorbents the energies of adsorption exceed heats of condensation of corresponding liquids, the critical pressures of 2DC are always less then the critical pressures of three-dimensional condensation of liquids, that is, then the saturation pressures, $p_c \ll p_s$. For example, for bulk graphite the ratio $p_s/p_c$ for a number of adsorbates falls in the range from 40 to 2000.[22] Because the adsorption energy is the sum of interactions with all atoms of the body, decreasing the body dimension leads to a decline of energy of interactions and an increase of $p_c$; it has been shown that for the finest graphite-like nanoparticles $p_c$ approaches the saturation pressure $p_s$ of the corresponding liquid.[38]

**5.2 Adsorption in porous media.** Consider adsorption in cylindrical capillaries with the diameters in excess of the several LD diameters of adsorbate. Condensation in such pores occurs at relatively high pressures, so that before condensation the cylindrical surface proved to have been covered by a liquid adsorption film due to pre-adsorption. The cylindrical pore covered by adsorption films can be considered as a liquid cylinder, since just the liquid film makes the main contribution to the LJ interactions of adsorbate with the wall. In pores, as well as on the flat surface, adsorption also begins to occur on the capillary walls and the 2DC initiates at $p_c$. Let us suppose that the layer-by-layer sequential adsorption occurs inside the cylinder. In this case, a formation of the new adsorption layer inside the cylinder would lead to diminishing the cylinder free space (cylinder radius) and, according to eqs 32 and 34, to the enhancement of molecular interactions. It would mean that each new adsorption layer must be formed at a lesser critical pressure of 2DC then the preceding one. The only way to resolve this paradox is to admit that, as the 2DC is occurring on the pore wall, a point is reached when the adsorption process is energetically as favorable for an adsorbate molecule to exist in the residual space as it is to complete the monolayer coverage; at this point, the 2DC begins to evolve in volume filling of a residual space at the same critical pressure of 2DC. [38, 39]

Of special interest is adsorption in slit-like pores (Figure 15).

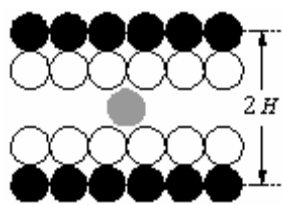

**Figure 15.** Schematic illustration of slit-like micropores of active carbons. The black disks and the circles represent carbon atoms of the walls and the monolayer adsorption films on the walls. The grey disk depicts the molecule of the third adsorption layer. The pore width is equal to $2H$, where $H$ is the pore half-width.

The theoretical treatment of adsorption in slit-shaped micropores proceeds from the approach initiated by Polanyi as long as 1932[40] and shortly afterwards by de Boer and Custers[41] who have assumed that the energy of gas-solid interactions is enhanced due to overlapping of the field of forces of neighboring walls in narrow pores. We shall call a micropore that can accommodate $n$-layers of molecules along its diameter as $n$-layer micropore. It is evident that a 2DC on the wall of a one-layer micropore is equivalent to the volume filling of the slit. In the case of two-layer micropores, due to the symmetry of walls, a 2DC occurs simultaneously on both walls and leads to volume filling of a void space. For three-layer micropores (Figure 15), although molecules of the middle layer are at a greater distance from solid, they find themselves in the strong field formed by adsorption monolayers, which overshadows a decrease of wall attractions.[42, 43] In this case again a point is reached when the 2DC begins to evolve in volume filling of a residual space at the same critical pressure. If $n \geq 4$, widths of internal micropores exceed two molecular diameters and the effect of overlapping of potentials practically vanishes; as a result, for micropores with $n \geq 4$, the adsorption potential in the residual space is less then that near the solid walls and the 2DC terminates on the pore walls. Afresh we come to a conclusion that the equilibrium pressure of pore filling is the critical pressure of 2DC on the pore wall; but in contrast to cylinders, for open slit-like pores there is a limit pore thickness above which a volume condensation does not occur.

**5.3 Main Equation in the case of Adsorption in Porous Media.** If one takes the critical pressure of 2DC and the energy of adsorption on the homogenous flat surface as the reference parameters for calculating equilibrium pressures over a pore formed by the same homogeneous surface, the physical reasoning again leads to the main equation 6. In this case, $\Delta(\Delta H)$ is the difference of adsorption energies in the pore and on the flat surface. Due to the importance of the equation, we shall derive it also in the framework of the statistical mechanics.[43]

It is convenient to consider the adsorbed film as a distinct phase with a known volume $V^{(s)}$, containing a known number of molecules, $N^{(s)}$, at a fixed temperature $T$. A superscript $s$ denotes that the thermodynamic properties relate to the adsorbed phase. The thermodynamic properties of the system are calculated from the canonical partition function, $Q(N,V,T)$:[26]

$$Q(N,V,T) = \frac{Z_N^{(s)}}{N! \Lambda^{3N}} \qquad (36)$$

where $\Lambda$ is the thermal deBroglie wavelength and $Z_N^{(s)}$ is the configurational integral:[26]

$$Z_N^{(s)} = \int_V \exp[-\left(\sum_{i=1}^{N} u_s(r_i) + \sum_{1 \leq i < j \leq N} u(r_{ij})\right)] d\mathbf{r}_1 d\mathbf{r}_2 .... d\mathbf{r}_N \qquad (37)$$

Here the potential energy of the $N$ molecules is taken to be the sum of the gas-solid energies, $u_s(r_i)$, and the mutual interaction energies of the fluid molecules with one another, $u(r_{ij})$. The vectors $r_i$ and $r_{ij}$ specify all variables relevant to the location of molecules $i$ and to the location of the pair $ij$. In our treatment, we retain definitions adopted by Steele.[26] An isotherm equation is derived from the general relation:

$$\ln\left\{\frac{p}{N^{(s)}kT}\right\} = -\left(\frac{\partial \ln Z_N^{(s)}}{\partial N}\right)_{T,A,N=N^{(s)}} \qquad (38)$$

where $A$ is the adsorbate area. A usual approach to derive two-dimensional expressions from general equations 36-38 involving molecules moving over a three-dimensional potential surface is to split the vector $r$ into two components ($\tau$, $z$), where vector $\tau$ is parallel to the surface and $z$-component of $r$ is the gas-surface separation, which is perpendicular to the surface, and then to assume that $u_s$ can be written as a separable function of $z_i$ and $\tau_i$[26]

$$u_s(\mathbf{r}_i) = f(\mathbf{\tau}_i) + \varepsilon(z_i) \qquad (39)$$

It has been shown that, if one introduces the quantity

$$Z_N^{(2D)} = \frac{Z_N^{(s)}}{(Z_1^{(s)})^N} \qquad (40)$$

and expands the Boltzmann factor for fluid-fluid interactions into a Tailor series centered about the well depth at $z_{min}$, then $Z_N^{(2D)}$ becomes:[26]

$$Z_N^{(2D)} \approx \frac{\int_A \exp[-\left(\sum_{i=1}^{N} f(\boldsymbol{\tau}_i) + \sum_{1 \leq i < j \leq N} u(\boldsymbol{\tau}_{ij})\right)] d\boldsymbol{\tau}_1 d\boldsymbol{\tau}_2 ...., d\boldsymbol{\tau}_N}{\int_A \exp[-\left(\sum_{i=1}^{N} f(\boldsymbol{\tau}_i)/kT\right)] d\boldsymbol{\tau}_1 d\boldsymbol{\tau}_2 ...., d\boldsymbol{\tau}_N} \quad (41)$$

Compare adsorption on two surfaces: (i) on the surface of a pore and (ii) on the reference flat surface. For example, micropore walls of activated carbons (AC) are similar to the basal graphite plane and it is natural to take the latter as a reference surface. Parameters related to the reference surface will be further supplied by a superscript "*ref*", whereas the absence of the superscript will refer to adsorption on surface of micropores. Keeping in mind eq. 40, equation 38 gives for adsorption on the reference surface:

$$\ln\left\{\frac{p^{ref}}{N^{(s)}kT}\right\} = -\left(\frac{\partial \ln\left[\left(Z_1^{(s),ref}\right)^N Z_N^{(2D),ref}\right]}{\partial N}\right)_{T,A,N=N^{(s)}} \quad (42)$$

and for adsorption on the pore surface

$$\ln\left\{\frac{p}{N^{(s)}kT}\right\} = -\left(\frac{\partial \ln\left[\left(Z_1^{(s)}\right)^N Z_N^{(2D)}\right]}{\partial N}\right)_{T,A,N=N^{(s)}} \quad (43)$$

Because $Z_N^{(2D)}$ and $Z_N^{(2D)ref}$ depend only on $\tau$-components and both surfaces have the same arrangement of adsorption sites

$$Z_N^{(2D)} = Z_N^{(2D),ref} \quad (44)$$

when $T$, $N_s$ and $A$ are held constant (as would be natural for constant amount adsorbed on equivalent substrates of the same areas). The Polanyi-de Boer-Custers[40, 41] assumption leads to

$$Z_1^{(s)} = \exp\left(-\frac{\Delta \varepsilon^*}{kT}\right) Z_1^{(s),ref} \quad (45)$$

where $\Delta\varepsilon^*$ is the excess gas-solid energy on the pore surface with respect to that on the reference surface. Subtracting eq. 42 from eq. 43 and keeping in mind eqs. 44 and 45, after some algebra we come to

$$\ln\frac{p}{p^{ref}} = \frac{\Delta\varepsilon^*}{kT} \qquad (46)$$

Equation 46 relates the equilibrium pressure of 2DC on the pore walls $p$ to that on the reference surface. As has been earlier shown, $p$ is the filling pressure of the individual micropore with the excess adsorption energy $\Delta\varepsilon^*$. One may see that eq 46 coincides with the main equation 6 and $\Delta(\Delta H) = \Delta\varepsilon^*$.

**5.4 Effect of the pore width on $\Delta(\Delta H) \equiv \Delta\varepsilon^*$.** It is clear that the energy of adsorption is the sum of energies of hydrogen bonds, dipole-dipole and the London dispersion interactions, depending upon the chemical nature of adsorbate and adsorbent. But the difference of the adsorption energies on two adsorbents with the identical arrangement of surfaces, just as in the case of autoadsorption on liquid bodies, is determined mainly by the difference of the dispersion interactions and may be calculated as for the LD interactions.

Calculation of $\Delta(\Delta H) \equiv \Delta\varepsilon^*$ for pores with "liquid" walls (cylinder coved by a liquid film) was given above. Consider now $\Delta\varepsilon^*$ in the case of adsorption on the bare walls (without films) of slit-like micropores of activated carbons. It is generally accepted that the pore walls of carbons are formed by graphite-like structures. The energy of adsorption on the reference graphite surface may be either taken from the literature or calculated proceeding from a dominant contribution of the dispersion interactions. The energy of interactions of a molecule with the individual graphite plane as a function of its separation, $z$, from the plane, $\varepsilon(z)$, is given as follows[44, 45]

$$\varepsilon(z) = \varepsilon^* \frac{10}{3}\left[\frac{1}{5}\left(\frac{1}{z}\right)^{10} - \frac{1}{2}\left(\frac{1}{z}\right)^{4}\right] \qquad (47)$$

where $\varepsilon^*$ is the energy of adsorption, that is, the value of $\varepsilon(z)$ at the equilibrium separation of the molecule from the plane, $z^*$. The energy of adsorption in a pore, $\varepsilon^*_p$, with a half-width $H$ is the sum of interactions with the both parallel walls: [45]

$$\varepsilon^*_p(2H, z^*) = \varepsilon^*(z^*) + \varepsilon^*(2H - z^*) \qquad (48)$$

where $(2H-z^*)$ is the distance of the molecule from the second wall. Figure 16 demonstrates the effect of widths of slit-like pores between graphite-like walls of activated carbons on (1) the excess energy of benzene adsorption in pores $\Delta\varepsilon^*$ in respect to that on the separated graphite surface and (2) on the equilibrium pressures of the benzene vapor calculated by eq 46. The adsorption energy on the non-porous graphite is equal to 40 kJ/mol and it is doubled in the slit that can accommodate only one molecule along its width (such a pore has a half-width of unity in Figure 16).

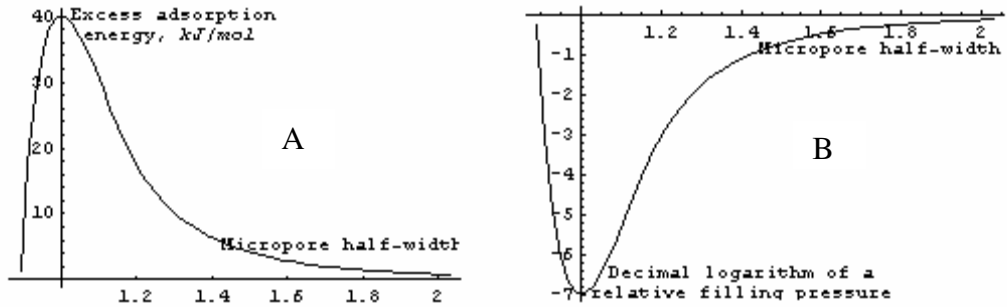

**Figure 16.** **(A)** Effect of a pore half-width, $H$, on the excess energy of benzene adsorption in slit-like carbon micropores. The excess energy is equal to the difference between the energy of adsorption in a micropore and that on the planar graphite surface. **(B)** Effect of the micropore reduced half-width on the decimal logarithm of a relative filling pressure, $\text{Log}_{10}(p/p_s)$. For micropore with $H \approx 1$, filling pressure of benzene reduces by factor $\approx$ ten millions.

The filling pressures of the individual micropores are given by eq 46. Just as in capillaries, the enhancement of adsorption fields in narrow micropores, leads to reduction of equilibrium pressure; but if in "liquid" capillaries with $r \approx 3$ it diminishes to $\approx 0.1 p_s$, whereas in a slitlike micropore of active carbon that can accommodate only one molecule along the pore width the pressure is reduced by factor up to $10^5$ (nitrogen, 78 K) or $10^7$ (benzene, 293 K). (Figure 16, right).

## 6. Experimental Verification

In the case of droplets and capillary condensation, the main equation competes with the famous Kelvin equation. For the spherical droplets with diameters more then 1 micron, for which there are direct experimental measurements, both equations lead to the close results.[22] Nevertheless, for the finest nanocapillaries the equations predict diverse results and a comparison with experiments demonstrates that an accurate description is given only by the main equation. According to the Kelvin equation,[25] the equilibrium pressure $p$ over a concave meniscus in capillaries is given by

$$RT \ln \frac{p}{p_s} = -\gamma V_m \left( \frac{1}{r_1} + \frac{1}{r_2} \right) \qquad (49)$$

where $r_1$ and $r_2$ are the radii of curvature and $V_m$ is the liquid molar volume.

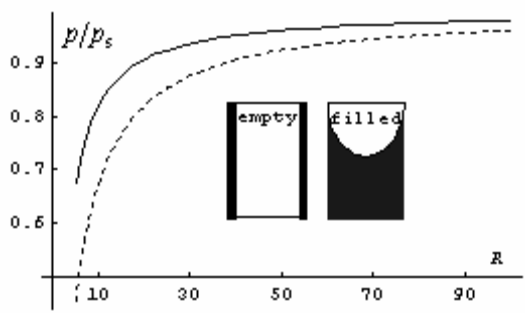

**Figure 17.** The Kelvin equation. Effect of cylinder radii $R$ on equilibrium pressures $p/p_s$ of water vapor for adsorption (solid line) and desorption (dashed line) at 293 K. **Insert**. Condensation occurs in the empty capillary when its wall is covered by the pre-adsorbed film (bold line) which takes a form of the cylinder. Desorption occurs from the filled cylinder

where the liquid surface has a hemispherical shape.

The characteristic feature of condensation in capillaries is a hysteresis loop. It is believed that the loop takes its origin from different forms of menisci in filled and empty capillaries (Insert. Figure 17). It is seen that condensation occurs in a cylindrical meniscus, whereas evaporation (desorption) takes place from a hemispherical ones.[25, 26] For the hemispherical meniscus $r=r_1=r_2$, where $r$ is the radius of capillary; for the cylindrical meniscus, $r=r_1$ and $r_2=\infty$. Substituting these values in eq 47 one obtains for the equilibrium pressures of desorption, $p_{des}$, and adsorption, $p_{ad}$:

$$RT \ln \frac{p_{des}}{p_s} = -\frac{2\gamma V_m}{r} \quad (50)$$

and

$$RT \ln \frac{p_{ad}}{p_s} = -\frac{\gamma V_m}{r} \quad (51)$$

It is seen from these equations that $p_{des}$ and $p_{ad}$ do not coincide, except the trivial case at $r=\infty$. Hence, the Kelvin equation predicts a divergence of hysteresis branches (Figure 17). Nevertheless, none of the porous systems conform to this model: thousands of experiments with a variety of adsorbents and adsorbates provide evidence that the capillary branches converge. Although many artificial assumptions[25, 26] were introduced, the quantitative explanations of peculiarities of hysteresis loops has not been given yet. It is generally accepted to use only the desorption branch for calculating mesopore radii by eq 50 and just to ignore eq 51. In contrast to the Kelvin equation, the main equation correctly describes the hysteresis loops (Figure 18).

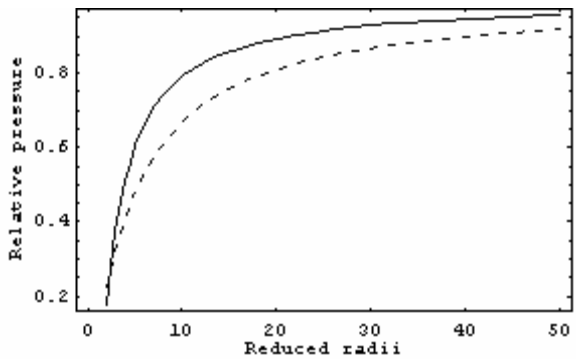

**Figure 18.** The new model. Effect of reduced radii on equilibrium pressures for water adsorption (solid line) and desorption (dashed line) at 293 K.

For water, the beginning of the loops is at $p/p_s \approx 0.35$ (Figure 18) that is close to the experimental data. The similar results were obtained for other adsorbates (nitrogen, argon, benzene).

The most famous attempts to verify the Kelvin equation were undertaken by Shereshefsky and co-worker in the course of 44 years (1928-1972).[46-50] They studied equilibrium in conical capillaries and found that the equilibrium radii of liquid columns in the capillaries were always greater then the values calculated from the Kelvin equation. Adamson[25] points out that the situation with the experimental verification of the Kelvin equation is still conflicting; whereas Everett, Haynes, and McElroy[51] in their conclusion remark draw attention to the fact that the Kelvin equation still lacks experimental verification! The data of Shereshefsky raised hot debate and have remained unexplained for decades and only the main equation proved to be in agreement with the Shereshefsky experiments.[33]

The main equation has been generalized to the condensation in the system of pores of random sizes. An isotherm of adsorption in such a system may also be viewed as a "stairway" (see Insert to Figure 14), the height of each jump and its position on the axes of pressure being determined by the volume of the individual pores, their quantities, and the energy of adsorption in the pore. Such an approach to adsorption has been realized proceeding from the hypothetical normal law of distribution of pores over their dimensions; it correctly describes the experimental isotherms,[43] provides calculations of pore size distributions and specific surface areas,[43] and allows predicting the adsorption isotherms.[43] A general picture of adsorption is in a good agreement with the results of computer simulations; the calculated values of micropore filling pressures proved to be in the quantitative agreement with those predicted by density functional theory. The model provides the rational foundation for the Dubinin equations,[52, 53] it bridges a gap between theories of adsorption on the surface and into the volume, and may be easily extended to adsorption on heterogeneous flat surfaces; it is particularly suitable for discussion of contribution of adsorption to adhesion and particle agglomeration.[54]

## 7. Further Discussion

The essence of the developed method is the following. The phase transition of the equilibrium system with well-established properties is taken as a reference process to search for the parameters of phase transition in the closely related, perturbed equilibrium system of interest. What physical meaning is attributed to the notion of "closely related systems"? Both systems must contain the same component in the vapor and liquid phases; otherwise, the variation of enthalpies of evaporations, $\Delta(\Delta H)=\Delta H_{ref}-\Delta H_{int}$, where the subscript *int* relates to the system of interest, has no physical significance ($\Delta(\Delta H)$ between different compounds is unknown in principle). The problem, the answer to which is sought for, may be formulated as follows: "How must the latent heat of the reference system be changed due to the perturbation to produce the observed variations in the equilibrium pressures?" In our treatment, the dissolution of a solute, surface bending, and external adsorption field are considered as the perturbing actions on the reference systems and the changes of enthalpies originate from the perturbations.

The main equation provides an account for the distinct perturbing effects in the framework of the united theory. Although the parent CE is a thermodynamically exact equation that contains no approximations, the main equation, as well as the Clausius-Clapeyron equation, contains several approximations made in the course of derivations: it is assumed that (1) the vapor phase behaves as ideal gas and (2) the variation of enthalpy $\Delta(\Delta H) = \Delta H_{ref} - \Delta H_{int}$ is independent of temperature. Beside these approximations, there is one more problem related to the accuracy of the main equation that is worth clearing up.

The comparable systems are in equilibrium at a given temperature and pressure when $\Delta G = \Delta H - T\Delta S = 0$, that is:

$$\Delta H_{ref} = T\Delta S_{ref} \qquad (52)$$

and

$$\Delta H_{int} = T\Delta S_{int} \qquad (53)$$

Subtracting eq 52 from eq 53, one has

$$\Delta(\Delta H) = T\Delta(\Delta S) \qquad (54)$$

where $\Delta(\Delta S) = \Delta S_{int} - \Delta S_{ref}$ is the difference of entropies of two phase transitions. These latter values depend not only upon the chemical nature of substances, but also upon such physical parameters as temperature and equilibrium vapor pressure. For a given temperature

$$\Delta(\Delta S) = [S^V(p_{int}) - S^L_{int}(p_{int})] - [S^V(p_{ref}) - S^L_{ref}(p_{ref})] \qquad (55)$$

Here superscript *V* and *L* refers to the vapor and liquid phases, respectively; symbols ($p_{ref}$) and ($p_{int}$) point that the values of entropy are taken at the corresponding equilibrium pressures. By regrouping the terms, we have

$$\Delta(\Delta S) = [S^V(p_{int}) - S^V(p_{ref})] - [S^L_{int}(p_{int}) - S^L_{ref}(p_{ref})] \qquad (56)$$

One may see that $\Delta(\Delta S)$ is combined from the changes of entropies of vapor, $\Delta S^V = [S^V(p_{int}) - S^V(p_{ref})]$, and the liquid phases, $\Delta S^L = [S^L_{int}(p_{int}) - S^L_{ref}(p_{ref})]$. The change of entropy of vapor originates from the difference of the equilibrium pressure; for the ideal gas the change of molar entropy is given as[1]

$$[S^V(p_{int}) - S^V(p_{ref})] = R \ln \frac{p_s}{p} \qquad (57)$$

Note that we return in eq 57 to the former definitions: $p_{ref} \equiv p_s$ and $p_{int} \equiv p$. If one assumes that $\Delta S^L = [S^L_{int}(p_{int}) - S^L_{ref}(p_{ref})]$ is close to zero, then eq 56 transform into

$$\Delta(\Delta S) \approx R \ln \frac{p_s}{p} \qquad (58)$$

Substituting this expression in eq 54, we obtain the main equation

$$\Delta(\Delta H) = -RT \ln \frac{p}{p_s}$$

From here, one may see that the equation is accurate if the difference of entropies of the liquid components taken at their equilibrium pressures is close to zero. For example, for aqueous solution of NaCl eq 58 is valid at such concentrations of NaCl for which entropy of liquid water at $p_s$ is close to the partial entropy of water in the solution at $p<p_s$; for adsorption in porous media, the condition 58 keeps hold if entropy of adsorption film on the nonporous surface is close to entropy of adsorbate in the pore. Therefore, the accuracy of the model is limited by eq 58; due to the absence of the reliable data, the further studies are needed to clarify the accuracy of the model for each specific application.

## 8. Conclusions

The alternative approach to the displacement of phase equilibrium is developed proceeding from the Clapeyron equation. The central position in the theory takes the variation of the enthalpy of evaporation originated from the change of energy of interactions due to the physicochemical perturbation of the system. The main equation, which establishes the correlation between $\Delta(\Delta H)$ and the equilibrium pressure, provides the unified method for studying a host of phenomena. The value of

Δ(Δ$H$) has a clear physical meaning and gives a new insight into our understanding of the phenomena, the correlations between which probably have earlier escaped notice.

**Acknowledgments**

I would like to thank Dr. Arkady Rutkovskiy, MD, PhD from the University of Oslo for his help in preparation of this manuscript.